# Higher-precision radial velocity measurements with the SOPHIE spectrograph using octagonal-section fibers

S. Perruchot[a], F. Bouchy[b,a], B. Chazelas[c], R.F. Díaz[b,a], G. Hébrard[b,a], K. Arnaud[d], L. Arnold[a], G. Avila[e], X. Delfosse[f], I. Boisse[b], G. Moreaux[g], , F. Pepe[c], Y. Richaud[a], A. Santerne[g], R. Sottile[a], D. Tezier[c]

[a]CNRS- Observatoire de Haute-Provence, F-04870 St-Michel-l'Observatoire ;
[b]CNRS- Institut d'Astrophysique de Paris, 98bis bd Arago, F-75014 Paris ;
[c]Observatoire de Genève, Université de Genève, 51 Chemin des Maillettes, 1290 Sauverny, Switzerland ;
[d]CNRS – Centre de Physique des Particules de Marseille, 163, avenue de Luminy, Case 902, F-13288 Marseille cedex 09;
[e]European Southern Observatory, Karl-Schwarzschild-Strasse 2, D-85748 Garching bei Muenchen.
[f]CNRS- Institut de Planétologie et d'Astrophysique de grenoble, BP53, F-38041 Grenoble Cedex 09;
[g]CNRS- Laboratoire d'Astrophysique de Marseille, Université d'Aix-Marseille & CNRS, 38, rue Frédéric Joliot-Curie F-13388 Marseille cedex 13;

**ABSTRACT**

High-precision spectrographs play a key role in exoplanet searches using the radial velocity technique. But at the accuracy level of 1 m.s$^{-1}$, required for super-Earth characterization, stability of fiber-fed spectrograph performance is crucial considering variable observing conditions such as seeing, guiding and centering errors and, telescope vignetting. In fiber-fed spectrographs such as HARPS or SOPHIE, the fiber link scrambling properties are one of the main issues. Both the stability of the fiber near-field uniformity at the spectrograph entrance and of the far-field illumination on the echelle grating (pupil) are critical for high-precision radial velocity measurements due to the spectrograph geometrical field and aperture aberrations. We conducted tests on the SOPHIE spectrograph at the 1.93-m OHP telescope to measure the instrument sensitivity to the fiber link light feeding conditions: star decentering, telescope vignetting by the dome,and defocussing.

To significantly improve on current precision, we designed a fiber link modification considering the spectrograph operational constraints. We have developed a new link which includes a piece of octagonal-section fiber, having good scrambling properties, lying inside the former circular-section fiber, and we tested the concept on a bench to characterize near-field and far-field scrambling properties.

This modification has been implemented in spring 2011 on the SOPHIE spectrograph fibers and tested for the first time directly on the sky to demonstrate the gain compared to the previous fiber link. Scientific validation for exoplanet search and characterization has been conducted by observing standard stars.

## 1. INTRODUCTION

High-precision radial velocity measurements are so far the main technique for search and characterization of exoplanetary systems. Up to date, almost 90% of the known extrasolar planets have been detected or established and characterized (for transiting planets) using this method. The sensitivity of this technique continuously increases, opening the possibility of exploring the domain of low-mass planets down to a few Earth masses as illustrated by the discoveries made by the state-of-art spectrograph HARPS [1] with a radial velocity precision below 1 m.s$^{-1}$.

Far from being an old-fashioned technique, Doppler measurements will remain at the forefront of exoplanet science for the coming years thanks to their capability to explore the domain of low-mass planets down to Earth masses, to discover and characterize multiple planetary systems, to perform long-term surveys to find true Jupiter-like planets, to establish the planetary nature and to characterize the transiting candidates of photometric surveys. In the aim to reach Earth-type planets and also measure the expansion of the universe, new spectrographs in the visible are now under study with the goal to reach in the near future a precision of only few cm.s$^{-1}$ (ESPRESSO@VLT [2], CODEX@E-ELT [3]).

To reach such a level of accuracy, one must avoid a certain number of limitations. Among these, stellar noise (activity, pulsation, surface granulation), contamination by external sources (moon, close-by objects, background continuum), and changes in the spectrograph feed (e.g. atmospheric dispersion, turbulence, guiding and centering errors) [4]. Unlike the "Iodine cell" calibration technique, which reduces the available spectral range (500-630 nm) and the throughput of the spectrograph, the "simultaneous Thorium" calibration technique allows a larger spectral range (380-680 nm) without flux losses, but it requires two fibers, one for the stellar beam and the other for the calibration lamp used to monitor the instrumental drift, assuming that both fibers follow exactly the same path inside the spectrograph. Additional instrumental factors, such as CCD cosmetics, charge transfer inefficiency, wavelength calibration errors, thermal instability effects, should also be mentioned. For an exhaustive discussion of these limitations, we refer to [5], where the authors highlight the impact of the spectrograph illumination as one of the main limiting factors.

The high-precision fiber-fed SOPHIE spectrograph, operating at the 1.93-m telescope of the Haute-Provence Observatory since November 2006, reached a Doppler precision of about 5 m.s$^{-1}$, similar to other high-precision spectrographs involved in exoplanet surveys. SOPHIE plays a very efficient role in the search for northern extrasolar planets [6], as well as in the Doppler follow-up of photometric surveys for planetary transit searches like SuperWASP [7], HAT [8], CoRoT [9] and Kepler [10].

A radial velocity precision of 5 m.s$^{-1}$, although well adapted for detection and characterization of giant exoplanets, is far to be appropriate for the low-mass planet domain like Super-Earth and Neptune which requires the 1-2 m.s$^{-1}$ precision. We have found that SOPHIE is strongly limited by the insufficient scrambling gain of the fibers. The instrument set-up and the intrinsic properties of the fiber-link make the instrument very sensitive to observing conditions such as seeing, guiding and centering errors, and telescope vignetting (see section 2). To significantly improve the Doppler precision, we designed a fiber link modification taking into account spectrograph operational constraints. We have developed a new fiber link which includes a piece of octagonal-section fiber, having good scrambling properties, lying inside the former circular-section fiber, and we tested the concept on a bench to characterize near-field and far-field scrambling properties (see section 3). This modification has been implemented in June 2011 on the fiber links of the SOPHIE spectrograph and tested for the first time directly on the sky by observing standard stars to demonstrate the precision gain (see section 4).

## 2. SOPHIE AND ITS LIMITATIONS IN RADIAL VELOCITY

### 2.1 A brief description of the SOPHIE spectrograph

The SOPHIE architecture and design are fully described in [11]. See also Table 1 for a characteristics summary. The optical concept is a double-pass Schmidt echelle spectrograph installed in an isothermal environment associated with a high efficiency coupling fiber feeding system, including simultaneous wavelength calibration. Two observation channels are available: High Efficiency mode (HE) and High Resolution mode (HR), to obtain higher throughput or better radial velocity precision respectively. Each channel has two fibers: one for the target and the other for the sky or simultaneous calibration lamp exposure, the same as used in ELODIE [12]. Fibers have a diameter of 100 $\mu$m, leading to a resolution of 39 000 in HE mode. Resolution is increased to 75 000 in HR mode thanks to an exit slit (of width 40.5 $\mu$m). A double-scrambler (symmetrical doublets arranged to exchange object and pupil spaces) is included in the HR fiber link to homogenize the spectrograph entrance illumination.

Table 1. Summary of SOPHIE instrumental characteristics up to June 2011

| Instrumental parameter | Value |
|---|---|
| Fiber acceptance on the sky | 3 arcsec |
| Spectrograph working aperture | f/3.6 |
| Pupil diameter | 200 mm |
| Typical resolution power [HE / HR] | 39 000 / 75 000 |
| Wavelength domain | 387 nm – 694 nm |
| Number of spectral orders | 39 |
| Pixel sampling per FWHM [HE / HR] | 6.7 / 2.7 |
| Detector | 1 CCD EEV 2kx4k (61 x 31 mm) Pixel size 15 $\mu$m |
| Internal precision of the wavelength calibration | 1 m.s$^{-1}$ |
| Observed instrumental velocity error | 5 m.s$^{-1}$ |
| Efficiency at 390 nm [HE / HR] | 4.6% / 1.9% |
| Efficiency at 550 nm [HE / HR] | 10.4% / 4.3% |
| Efficiency at 690 nm [HE / HR] (including atmosphere, telescope, Cassegrain adapter, fibers, spectrograph and CCD) | 8.3% / 3.5% |
| S/N ratio per pixel at 550 nm | S/N = 100 in 1 hour on mag 11 (HE) S/N = 100 in 1 hour on mag 10 (HR) |

## 2.2 Radial velocity precision limitations

Fiber-fed spectrographs such as HARPS or SOPHIE have a radial velocity precision limited by several error sources as described in [5]. Among them, Doppler measurements are affected by atmospheric effects such as spectral dependent absorption or contamination, and chromatic flux variations mostly produced by the atmospheric dispersion combined with the varying seeing, as well as by varying atmospheric extinction. There are also instrumental effects whose amplitudes are linked to the architecture and design of the instrument: guiding, centering and focus stability, spectrograph illumination stability, instrument stability with regard to the environmental conditions, detector characteristics, wavelength calibration.

In the historical context of SOPHIE's development, its design was driven by the aim to significantly enhance (by a factor of 10) the instrument throughput compared to ELODIE, the previous spectrograph at the 1.93 m telescope, in addition to increased stability. This specification enforced a very compact design with few optical elements number and a large working aperture (f/3.6), which explains why SOPHIE geometric aberrations are large compared to those of HARPS. Images are affected by field aberrations (especially distortion and coma because the fibers are far from the collimator axis) and aperture aberrations (spherical and astigmatism). Thus variations of illumination at fiber link output, in near field or in far field, have critical consequences on spectrum stability on the CCD. For this reason, SOPHIE is more sensitive than HARPS to illumination variations in the input fiber due to seeing variations or guiding and centering errors as already described in [13].

# 3. SOPHIE FIBER LINK MODIFICATIONS

## 3.1 Octagonal fibers: a way to improve scrambling efficiency

Despite the very successful results obtained with SOPHIE since the end of 2006, a precision better than 5 m.s$^{-1}$ is required in the search for Super-Earth and Neptune-like exoplanets. As identified in previous section, improving the fiber scrambling properties before the spectrograph entrance is required to increase the precision obtainable with SOPHIE. As square and polygonal light pipes are a classical way to get more homogeneous illumination [14], such fibers have been considered for the next generation of precision radial velocity instruments as first proposed by P. Connes (1999 private communication). After promising simulations, Geneva Observatory , ESO and OHP teamed up to procure and characterize new fibers (octagonal or square sections from various manufacturers and dimensions), confirming the expected scrambling gain [15] [16].

Implementing octagonal fibers on the SOPHIE fiber links permits to test directly on the sky the increase in scrambling gain. Because of its very high sensitivity to variable illumination, SOPHIE is a good test bench for fiber-fed spectrographs to experiment new scrambling systems. To allow further understanding of the role played by the scrambling, the operation consisted in inserting a piece of octagonal-section fiber, lying inside the former circular-section fiber link, so that no other hardware is changed. The instrumental development consisted in two essential parts: validating the circular-octagonal-circular link in laboratory in terms of scrambling property (section 3.2) and implementing safely the new hardware (section 3.3).

## 3.2 Laboratory tests for fiber link losses and scrambling properties

Two properties had to be characterized in preliminary lab tests. First, the circular-octagonal-circular junctions should not affect too much the throughput of the spectrograph. Losses are a combination of effects: geometrical mismatch between circular and octagonal cores (~5%), decentering and misorientation at each junction, and Focal Ratio Degradation (FRD) due to constraints at the fibers terminations or some other effects. Second, the scrambling gain should be proved by comparison with a circular-only fiber link.

The circular-section fibers used in SOPHIE are Polymicro FVP100110125, with a core diameter of 100$\mu$m and a cladding diameter of 110$\mu$m. The octagonal-section fiber is a preform for an octagonal fiber of 70$\mu$m- core diameter from Ceramoptec; the preform octagon inner circle has a diameter of 100$\mu$m, the cladding is 187$\mu$m while the polyimide buffer is 210$\mu$m. We tried several techniques to join the octagonal fiber to the circular fiber. Core-to-core fusion splicing assures very good core alignment, but at a price of stress at the junction that produces FRD. Several tries on circular fibers gave inhomogeneous FRD results. We encountered many difficulties in splicing our octagonal fiber to the circular fibers essentially due to their very different cladding and buffer dimensions. We then decided to terminate this approach and to use FC to FC mating sleeves (available for instance in the Thorlabs® catalog). These mating components contain a centering ring, and the contact between fibers is maintained by a spring system inside the FC connectors. This connection achieves very good performance (excellent centering and very low FRD), possibly enhanced with index-matching liquid to reduce Fresnel losses in case of non perfect optical core contact. See on Figure 2 (left) the measurements achieved with the experimental set-up described in Figure 1 for a circular-circular junction (pink curve) compared to a single circular fiber (blue curve).

Metallic FC connectors have been especially drilled to receive the octagonal fibers, but unfortunately the first set showed some obvious decentering. Laboratory tests (set-up Figure 1) showed that the induced losses with the connection to circular FC fibers were acceptable in SOPHIE for a temporary implementation in order to validate their scrambling properties. As can be seen in Figure 2 (left), FRD is increased with a circular-octagonal-circular link (green curve) compared to circular-only connections. But we obtained the same FRD even in the fusion spliced case than in the FC to FC mating sleeve case (Figure 2 right side); it should be noticed that in both cases the octagonal fiber was well decentered. This FRD is not yet well-understood.

Scrambling properties were measured in the Geneva Observatory facility; see Figure 3 and [15] for a full description of the experimental set-up and measurement technique. The near field scrambling is measured using an image of the output of the fiber through a microscope objective. The beam at the fiber entrance is scanned across the core of the tested fiber assembly, simulating guiding errors in the telescope. The scrambling gain is defined as the ratio between maximum displacement of the barycenter of the near field illumination at the fiber output and the displacement of the spot at the fiber entrance.

We compared results between a SOPHIE-type circular fiber and an octagonal fiber that was fusion spliced inside a SOPHIE-type circular fiber (two splices). The Figure 4 illustrates the scrambling measurements. Table 2 summarizes the obtained scrambling gains. The use of an octagonal piece lying inside a circular fiber link is validated with a scrambling gain larger than a factor 10. We checked with a fusion spliced circular-section fiber assembly that FRD slightly increases the scrambling gain, even in near field. Furthermore one noticed that in a general way, the faster the beam, the lower the scrambling gain. This last property explains at least partially the lower scrambling gain of the f/3.6 SOPHIE spectrograph compared to HARPS (f/4) and ELODIE (f/5.6). There is also a slight scrambling gain in far field, identified as an FRD consequence.

All these laboratory results encourage us to implement octagonal fibers in the SOPHIE fiber links, expecting near-field scrambling gain without losing too much throughput (near 20%).

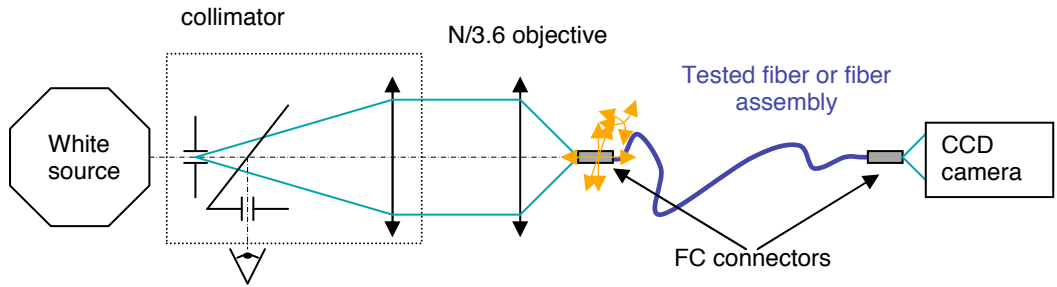

Figure 1. Experimental set-up in LAM (Marseille) for technical validation: images in intermediate space at fixed distance, fixed beam speed (f/3.6). Same set-up for every tested fiber assembly (octagonal fiber spliced to two circular fibers for instance). This allows relative FRD estimation, losses estimation and constraints detection by comparison to a reference fiber. The pinhole size can be chosen between an entire illumination of the fiber or half the fiber-diameter.

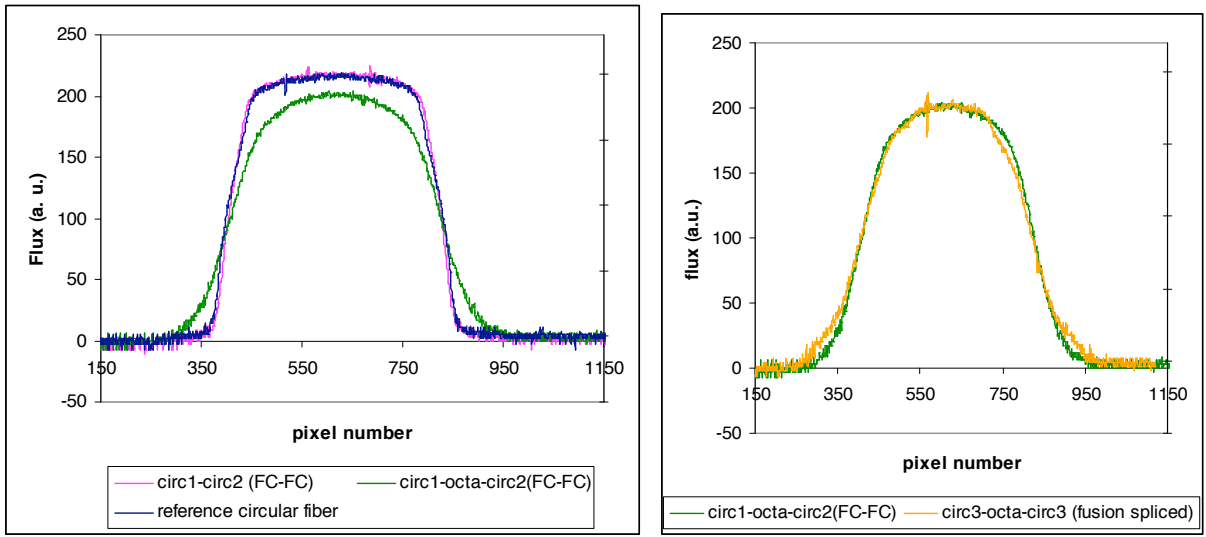

Figure 2. Cross sections of images obtained with the LAM set-up Figure 1. Left: comparison of links with FC to FC mating sleeves; note the FRD and the losses of the decentered octagonal link. Right: comparison between two circular-octagonal-circular fibers assemblies, both with decentered octagonal piece, the green curve with FC to FC mating sleeve connections, and the orange curve with fusion splice connections.

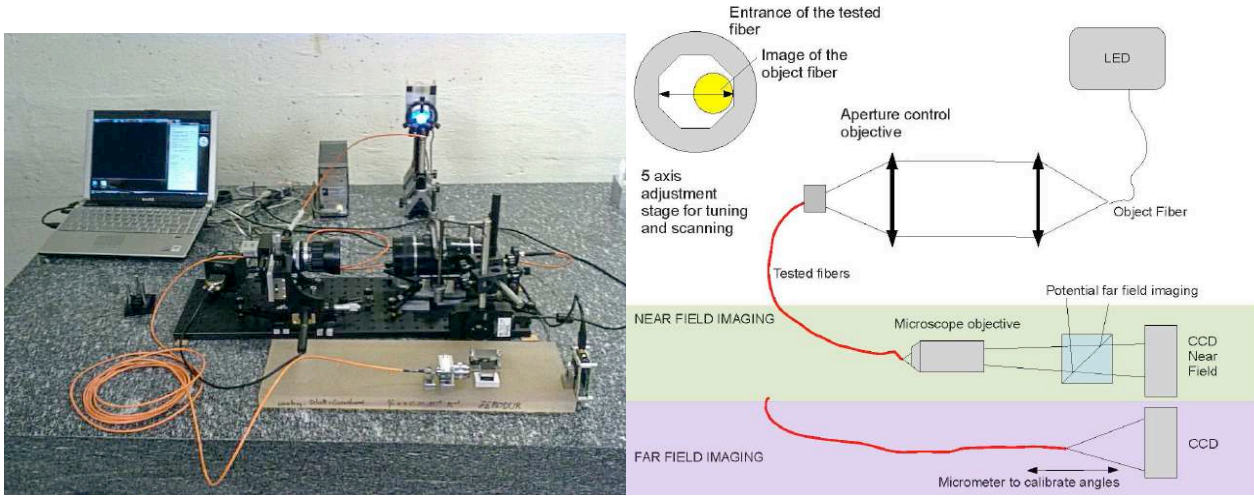

Figure 3. Geneva Observatory scrambling test bench. This test bench allows illuminating the tested fiber assembly in a controlled way (including the illumination spot diameter) and making pictures of near and far fields in a stable way in order to measure scrambling and FRD. Figures extracted from [15].

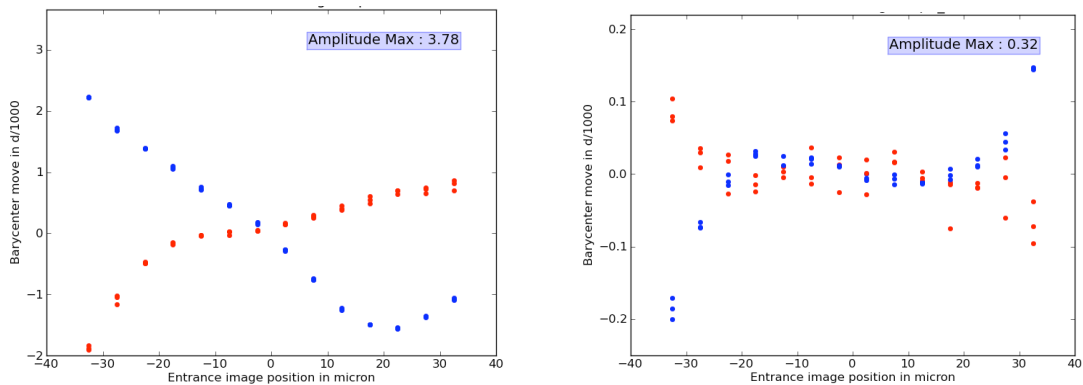

Figure 4. Near Field scrambling measurements with Figure 3 set-up. Left: for a FVP circular fiber 100 $\mu$m. Right: for the octagonal piece fusion spliced between two pieces of circular fiber 100 $\mu$m. The points represent the barycenter displacements of the output near field during a scan of a 35$\mu$m – diameter spot at N=2.3 on the core of the tested fiber assembly, in units of 1000$^{th}$ of fiber diameter. Red and Blue points are for the X and Y axis.

Table 2. Scrambling gain measured with Geneva bench for three fiber assemblies for 70-$\mu$m displacement of a 35$\mu$m diameter-spot.

|  | Circular fiber | Circ-circ-circ (fusion splices) | Circ-octa-circ (fusion splices) |
|---|---|---|---|
| Spot diameter 35$\mu$m, f/2.3 | 180 | 370 | >2200 |
| Spot diameter 35$\mu$m, f/4 | 240 | 730 | NA |

### 3.3 In situ implementation of octagonal fibers

Several possibilities are offered on the octagonal piece's localization on the HR fiber links: before, before and after or in the place of the optical double scrambler, the system that optically exchanges near field and far field as explained in [11]. Different results are expected as the octagonal fiber increases the scrambling gain only in the near field. Replacing the optical double scrambler has the advantage of reducing losses (about 20%) and attenuating guiding error effect and near-field seeing impact; the HE channel will work in this mode. Having an octagonal piece before the scrambler offers the advantage to attenuate guiding error effects and near-field seeing impact on the spectrograph pupil space, while the spectrograph slit is not affected in near-field. Adding a second octagonal piece after the double scrambler allows reducing far-field effects on the spectrograph slit, at the price of second octagonal-circular junction loss.

We benefit from the very good performance of FC-FC connections with circular fibers to prepare the implementation of all those possibilities and add FC connectors in all interesting places as shown in the lower part of Figure 5. We paid attention to have a minimal fiber length of 45 cm to allow correct azimuthal scrambling from the circular pipes. SOPHIE is a fully operational instrument, with more than 90% of nights a year dedicated on the 1.93m – telescope. Each instrumental modification has to be implemented in a short time to avoid a too large lack of observations in the scientific programs, and with a risk management plan to reduce failure possibilities to a bare minimum. Insertion of octagonal fiber pieces is localized near the spectrograph entrance, where no fiber movement happens, constraining to work in the isothermal small space available. As the chosen junction technique is to interconnect FC connectors, each initial fiber had to be cut, bonded to a FC connector and polished *in situ*, requiring protection of all the spectrograph optics. No local diagnostic except microscope viewing of the polished surface was available during technical intervention. As the SOPHIE fibers at f/3.6 are very sensitive to constraints in the termination, a great part of the operation success was based on the extensive experience of our technician in handling these fibers.

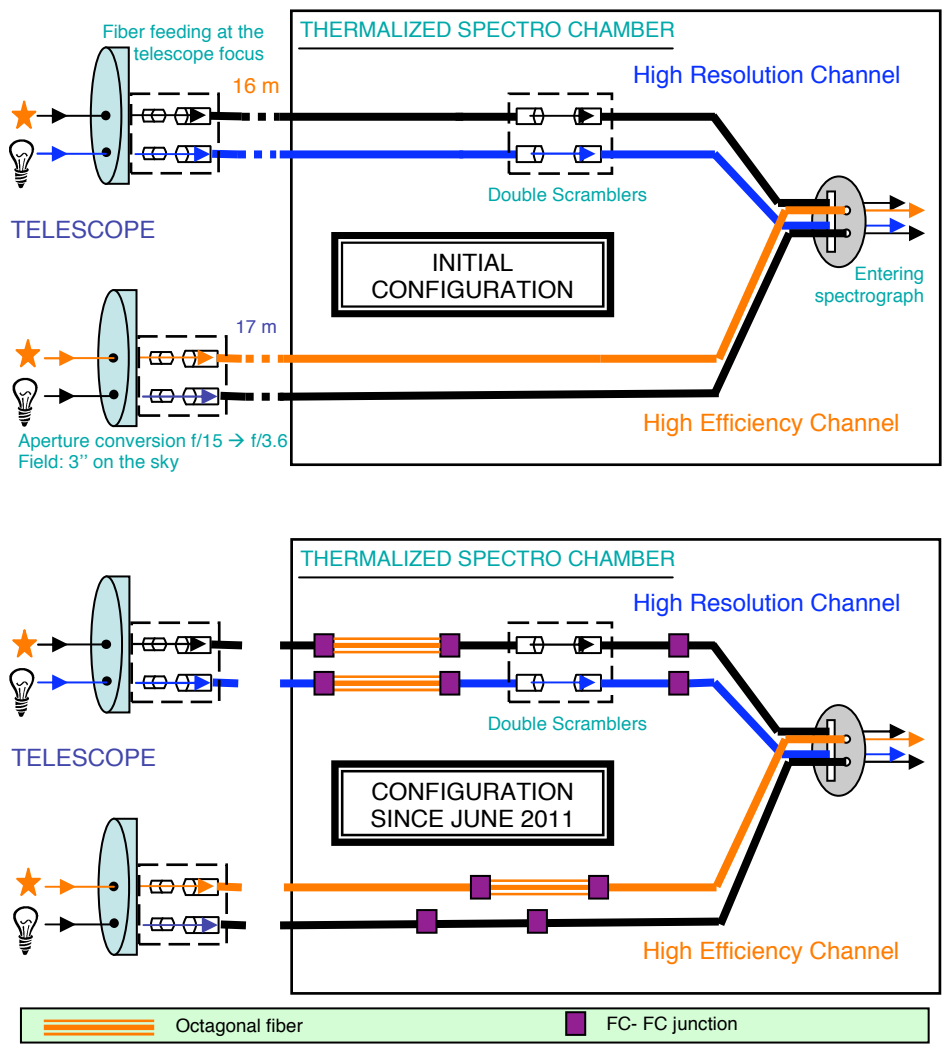

Figure 5. SOPHIE fiber links before (upper part) and after (lower part) the modification of June 2011: octagonal-section fiber insertion.

For the first implementation of the modified SOPHIE fiber links in June 2011, we had available 1.5-m octagonal pieces that were slightly decentered in their FC connectors, as explained in section 3.2. Nevertheless, loss estimates (~20%) led us to validate the use of three of them for a temporary setup: two of them in front of the double scrambler on the HR channel, and one on the target HE channel, the calibration link being less crucial in principle. This configuration allows testing two operational modes, one direct on HE channel, and one with space exchange in HR channel. When well-centered FC- connected octagonal fibers will be available, they will be implemented on each link, reducing losses, and other configurations could be tested on the HR channel.

Technical validation of the implementation was achieved by controlling the flux before and after the operation. This was done using a tungsten calibration lamp, allowing spectral loss control along the blaze with a precision better than 2-3%. Measured losses are -20%, -22% and -21% on HR star, HR calibration and HE star fiber link respectively, as estimated during laboratory tests considering geometrical and decentering losses on one hand, and FRD losses on the other hand.

## 4. NEW SOPHIE SPECTROGRAPH PERFORMANCE

We conducted systematic tests on the SOPHIE spectrograph at the 1.93-m OHP telescope to measure the instrument sensitivity to the fiber link light feeding conditions. These tests were performed before and after the new fiber links implementation.

### 4.1 Star decentering

We tested the sensitivity to guiding and centering effects on the fiber entrance with radial velocity sequences on standard stars. We fixed a given offset of the guiding system to impose a shift of the guiding set point to 1.0, 1.5 and 2.0 arcsec from the center of the fiber (3 arcsec diameter) in the four directions North, South, East and West. Figure 6 (Left) shows one sequence obtained with a seeing of about 1.8 arcsec using the HR mode with the initial configuration (showing RV variations reaching up to 30 m.s$^{-1}$ with a guiding shift of 2 arcsec) and with the new fiber links configuration. Figure 6 (Right) compares the RV change as a function of the position of the center of gravity of the star image at the fiber entrance for both modes (HE and HR). The sensitivity to the guiding decentering is clearly reduced by at least a factor 6 in both modes with the new fiber links. With the typical precision of the guiding system (0.2-0.3 arcsec), we now do not expect RV changes larger than 1 m.s$^{-1}$ due to guiding errors.

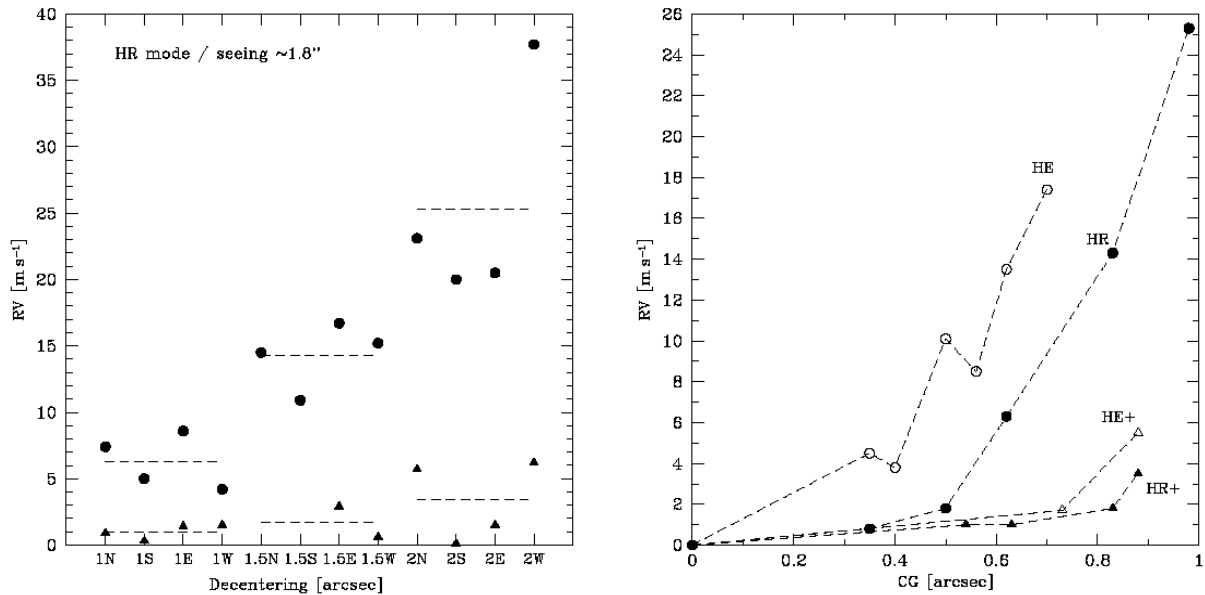

Figure 6. (Left) Sequence of radial velocity measurements with star decentering in HR mode with previous circular fiber links (black circle) and with the new links including octagonal fiber (black triangle). (Right) Radial velocity changes as a function of the position of the center of gravity (CG) of the star image at the fiber entrance in HE and HR modes with previous (circles) and new (triangles) fiber links configurations.

## 4.2 RV dispersion with standard RV stars

RV measurement precision was evaluated by sequences on the standard RV stars in HR mode, made before and after the implementation of the octagonal fibers. As shown on Figure 7, the new fiber links reduce the dispersion by a factor of about 6. Dispersion was 7.2 and 7.8 m.s$^{-1}$ with the previous configuration for HD109358 (mv=4.3) and HD185144 (mv=4.7) respectively. With the new configuration, we obtained precisions over 20 days of 1.8 m.s$^{-1}$ and 1.2 m.s$^{-1}$.

To go further, a seeing effect had been identified ([13]), illustrating RV variations due to seeing conditions. Figure 8 shows the RV obtained on standard stars as a function of the seeing. The seeing factor is computed as the flux per unit of exposure time monitored by the spectrograph. We see a clear anti-correlation of the RVs with this seeing factor for the previous fiber configuration with an amplitude of up to 25 m.s$^{-1}$. With the new octagonal fiber the seeing effect seems to be negligible or at least smaller than 2 m.s$^{-1}$.

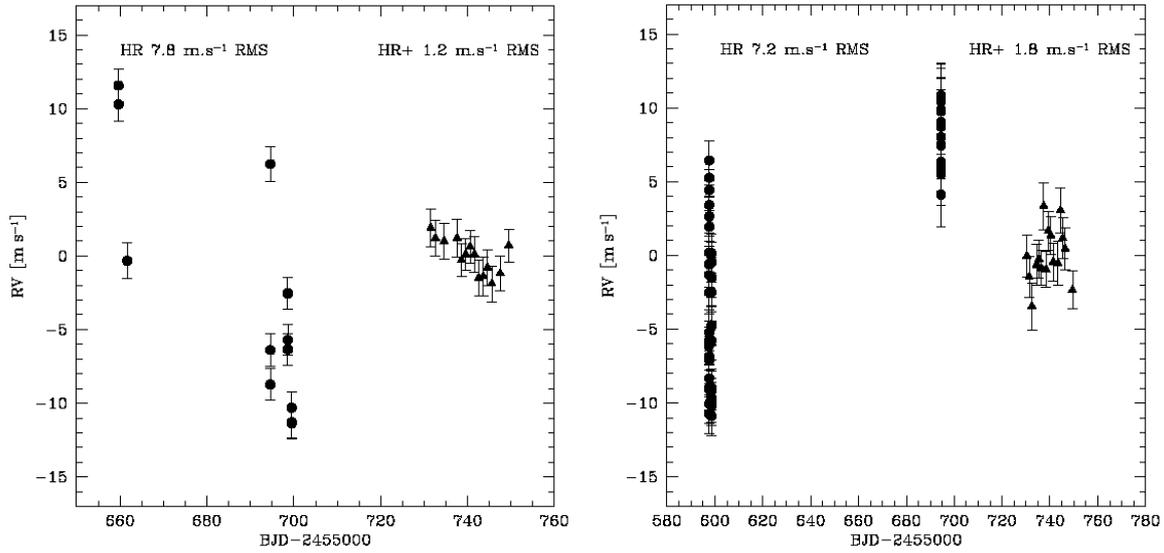

Figure 7. Radial velocity sequences made on the standard RV stars HD185144 (Left) and HD109358 (Right) before and after the octagonal fiber implementation (BJD = 2455720).

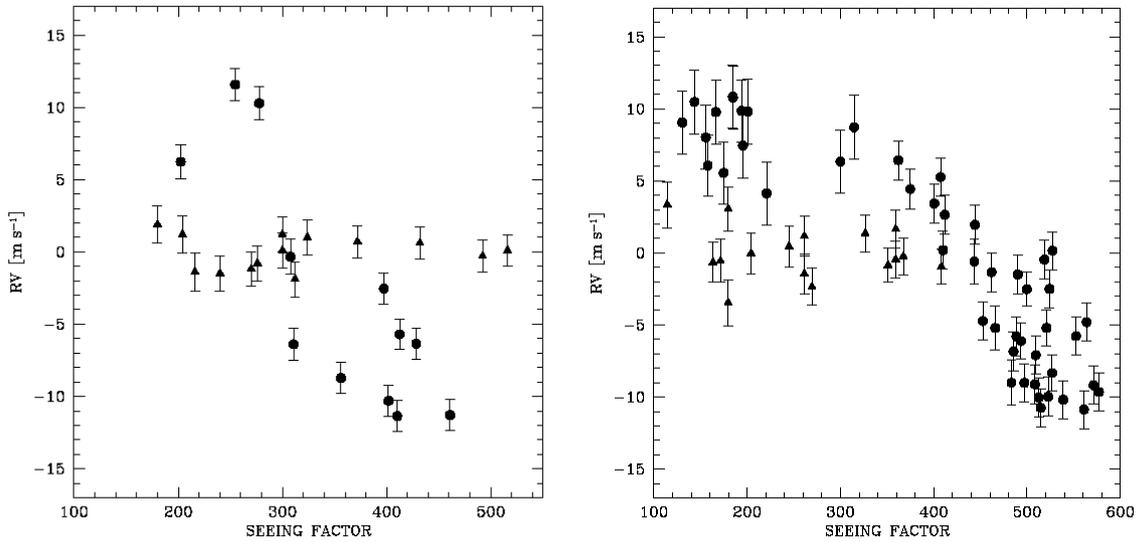

Figure 8. Radial velocity changes as function of the seeing factor showing a clear correlation when the previous configuration is used (circles). Triangles: use of the new configuration. Left: HD185144. Right: HD109358.

### 4.3 Defocussing and dome vignetting effects

We also tested the effect of a change in the far field of the fiber entrance illumination by defocussing the telescope or by vignetting of the telescope with the help of the dome.

RV changes up to +8 m.s$^{-1}$ and about -5 m.s$^{-1}$ in HR and HE modes respectively when about 40% of the telescope aperture is vignetted. We find the same amplitude using the new fiber link configuration as the previous one, as predicted since the octagonal fiber is not expected to affect the far field.

Testing the defocussing of the telescope is not equivalent to a seeing increase, but allows comparing the two fiber configurations with a test affecting both near and far field illumination. Increasing the apparent width of the star image from 2 arcsec up to 4-5 arcsec (controlled on the guiding camera) changed the corresponding RV in HE and HR modes up to -20 m.s$^{-1}$ and +10 m.s$^{-1}$ respectively. With the new fiber link configuration, the amplitude is slightly reduced to 12 m.s$^{-1}$ and 3.5 m.s$^{-1}$ respectively. This may confirm that defocussing affects probably both near and far field of the fiber illumination, but additional tests should be done to better characterize this effect.

### 4.4 Intrinsic calibration stability

We checked the intrinsic performances of the two SOPHIE fibers to measure the same instrumental drift. We computed the drift difference measured on each mode for both fibers using all the Thorium-Argon double calibration made during a full month before and after the implementation of the octagonal fibers. Figure 9 shows the dispersion of the drift difference. With the previous fiber link we found 2.27 m.s$^{-1}$ rms and 0.57 m.s$^{-1}$ for the HE and HR mode respectively, while with the new configuration we find 1.10 m.s$^{-1}$ and 0.22 m.s$^{-1}$. For the HR mode, the gain in relative stability is a factor of more than 2.5, and only 2 for HE mode where one fiber link only is equipped with an octagonal fiber.

For comparison, the typical daily drift of the spectrograph is 20-30 m.s$^{-1}$. This means that in HR mode the two fibers follow the spectrograph drift with a precision 100 times better. Note also that the photon noise on ThAr spectra is estimated to about 10 cm.s$^{-1}$.

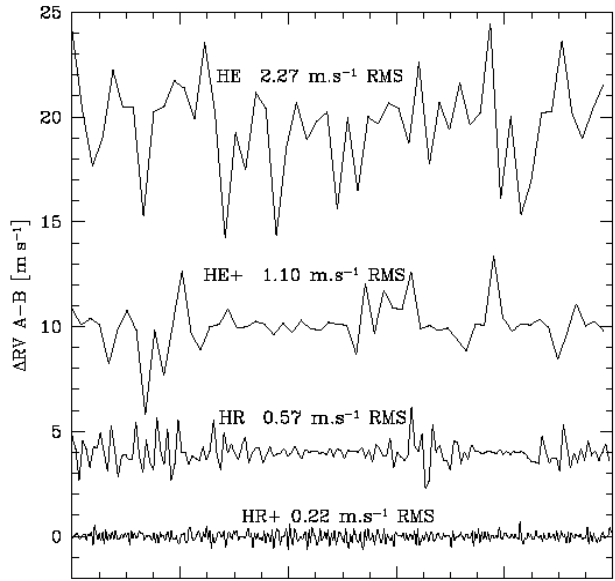

Figure 9. Radial velocity drift difference measured on fiber A (star channel) and B (calibration channel) for HE and HR mode before and after (+) the octagonal fibers implementation. The vertical scale is arbitrarily shifted to separate the four curves.

# 5. CONCLUSIONS AND PERSPECTIVES

We have successfully implemented octagonal-section fiber pieces in SOPHIE spectrograph fiber links using FC to FC mating sleeves. For the first time, the improvement by comparison with circular only- fiber links has been clearly demonstrated by tests directly on the sky with SOPHIE spectrograph. Scientific validation for exoplanet search and characterization has been started by observing standard radial velocity stars. The Doppler precision on a time scale of 20 days is better than 2 m.s$^{-1}$. Long term performance will be tested in the next months. We showed that the guiding, centering and seeing effects on the near field of the entrance fiber are reduced by more than a factor 6 thanks to the octagonal fiber and are now below 2 m.s$^{-1}$. The precision of the simultaneous drift measurement at 0.22 m.s$^{-1}$ in HR mode is now close to the photon noise of the Thorium-Argon spectra. The better precision achieved allows further investigation of other instrumental effects at the level of 1 m.s$^{-1}$.

We will implement new octagonal links as soon as more convenient ones will be delivered to improve throughput by reducing decentering effects and maybe FRD losses. The new fiber link arrangement allows also further configurations as adding an octagonal fiber after the optical double scrambler (to stabilize far-field illumination if necessary) or replacing the scrambler itself by a unique octagonal fiber. Results shall be exploited in the SOPHIE context as effects are spectrograph-dependent.

Octagonal-section fibers open a very significant improvement of fiber-fed spectrograph dedicated to high-precision radial velocity measurements with the possibility to reach the 1 m.s$^{-1}$ level. Future spectrographs now being developed like HARPS-North, SPIROU, and ESPRESSO will have benefit strongly from the use of such octagonal fiber link.

## ACKNOWLEDGEMENTS

The authors would like to thank all the staff of Haute-Provence Observatory for their contribution to the success of the SOPHIE upgrade and their support at the 1.93-m telescope. We wish to thank the Institut National des Sciences de l'Univers (INSU), the "Programme National de Planétologie" (PNP) of CNRS/INSU, the French National Research Agency (ANR-08-JCJC-0102-01) and the Swiss National Science Foundation for their continuous support to the SOPHIE project.

François Bouchy is grateful to Pierre Connes for all his useful advices on fiber scrambling.